\documentclass[twocolumn,aps,prd,10pt,superscriptaddress,longbibliography]{revtex4-1}
\usepackage{amsmath}
\usepackage{amssymb}
\usepackage{bm}
\usepackage{glossaries}
\usepackage{graphicx}
\usepackage{multirow}
\usepackage{fancyvrb}
\usepackage[linesnumbered,ruled]{algorithm2e}
\usepackage{siunitx}
\usepackage{subfigure}
\usepackage{threeparttable}
\usepackage{float}

\usepackage[usenames,dvipsnames,svgnames]{xcolor}
\usepackage{hyperref}
\hypersetup{
pdfnewwindow=true,      
colorlinks=true,        
linkcolor=Blue,         
citecolor=Blue,         
filecolor=Blue,         
urlcolor=Blue           
}

\newacronym{ace}{ACE}{atomic cluster expansion}
\newacronym{dft}{DFT}{density-functional theory}
\newacronym{eam}{EAM}{embedded-atom method}
\newacronym{ml}{ML}{machine learning}
\newacronym{ip}{IP}{interatomic potential}
\newacronym{dp}{DP}{deep potential}
\newacronym{gap}{GAP}{Gaussian approximation potential}
\newacronym{md}{MD}{molecular dynamics}
\newacronym{mlp}{MLP}{machine-learned potential}
\newacronym{mtp}{MTP}{momentum tensor potential}
\newacronym{nep}{NEP}{neuroevolution potential}
\newacronym{tabgap}{tabGAP}{tabulated Gaussian approximation potential}
\newacronym{snes}{SNES}{separable natural evolution strategy}
\newacronym{pc}{PC}{principal component}
\newacronym{2d}{2D}{two-dimensional}
\newacronym{rmse}{RMSE}{root-mean-square error}
\newacronym{zbl}{ZBL}{Ziegler-Biersack-Littmark}
\newacronym{hea}{HEA}{high-entropy alloy}
\newacronym{rhea}{RHEA}{refractory high-entropy alloy}
\newacronym{fp}{FP}{Frenkel pair}
\newacronym{nrt}{NRT-dpa}{Norgett-Robinson-Torrens displacements per atom}
\newacronym{arc}{arc-dpa}{athermal recombination corrected displacements per atom}
\newacronym{kp}{KP}{Kinchin and Pease}
\newacronym{bca}{BCA}{binary collision approximation}
\newacronym{dpa}{dpa}{displacements per atom}
\newacronym{sia}{SIA}{self-interstitial atom}
\newacronym{pka}{PKA}{primary knock-on atom}
\newacronym{bcc}{bcc}{body-centered cubic}
\newacronym{fcc}{fcc}{face-centered cubic}
\newacronym{w}{W}{tungsten}
\newacronym[longplural={threshold displacement energies}]{tde}{TDE}{threshold displacement energy}

\begin{document}

\title{Utilizing a machine-learned potential to explore enhanced radiation tolerance in the MoNbTaVW high-entropy alloy}

\author{Jiahui Liu}
\affiliation{Beijing Advanced Innovation Center for Materials Genome Engineering, University of Science and Technology Beijing, Beijing 100083, P. R. China}

\author{Jesper Byggm\"{a}star}
\email{jesper.byggmastar@helsinki.fi}
\affiliation{Department of Physics, P.O. Box 43, FI-00014 University of Helsinki, Finland}

\author{Zheyong Fan}
\affiliation{College of Physical Science and Technology, Bohai University, Jinzhou 121013, China}

\author{Bing Bai}
\affiliation{China Institute of Atomic Energy, Beijing, 102413, P. R. China}

\author{Ping Qian}
\email{qianping@ustb.edu.cn}
\affiliation{Beijing Advanced Innovation Center for Materials Genome Engineering, University of Science and Technology Beijing, Beijing 100083, P. R. China}

\author{Yanjing Su}
\email{yjsu@ustb.edu.cn}
\affiliation{Beijing Advanced Innovation Center for Materials Genome Engineering, University of Science and Technology Beijing, Beijing 100083, P. R. China}

\date{\today}

\begin{abstract}

High-entropy alloys (HEAs) based on tungsten (W) have emerged as promising candidates for plasma-facing components in future fusion reactors, owing to their excellent irradiation resistance. To achieve physically realistic descriptions of primary radiation damage in such multi-component materials, we propose extended damage models and trained an efficient machine-learned interatomic potential for the MoNbTaVW quinary system. From cascade simulations at primary knock-on atom (PKA) energies of 1–150 keV, we fitted an extended arc-dpa model for quantifying radiation damage in MoNbTaVW. Furthermore, we performed 50 cascade simulations at the recoil energy of 150 keV with 27.648 million atoms to investigate the effect of PKA types (Mo, Nb, Ta, V, W). The results show that subcascade splitting effectively suppresses interstitial cluster formation, which is a key mechanism for enhancing radiation resistance in HEAs. Our findings provide valuable insights into the radiation resistance mechanisms in refractory body-centered cubic alloys and highlight the potential of machine learning approaches in radiation damage research.

\end{abstract}

\maketitle

\section{Introduction}

\Glspl{hea} have attracted considerable attention for their outstanding material properties such as high strength, ductility, toughness and corrosion resistance~\cite{Senkov2010intermetallics, Miracle2017acta, Senkov2018jmr, George2019nrm, Coury2019acta, Senkov2019acta}.
Moreover, \gls{w}-based \glspl{rhea} exhibit outstanding radiation resistance~\cite{Waseem2021mcp, Zong2022nme, Wang2024jnm}. 
Recent studies show minimal radiation hardening and no signs of radiation-induced dislocation loops in nanocrystalline thin films of W-based \glspl{hea} even under high dose conditions~\cite{Atwani2019sciadv, Atwani2023nc}. 
Similarly, irradiated coarse-grained WTaCrV exhibits TEM-visible dislocation loops that are signifigantly smaller than in pure W~\cite{Atwani2023HEAM}. 
Despite these findings, the fundamental mechanisms underlying this irradiation resistance remain unclear, owing to the challenges in analyzing the defect generation and evolution mechanisms at the atomic level through experimental techniques.

The \gls{md} method is an effective tool for simulating displacement cascade processes, as well as defect generation, interaction, and migration behaviors at the atomic scale~\cite{Nordlund2019jnm, Deng2023jnm, Liu2023jnm, Li2024mtc, He2024jac, Guo2024msmse, Liu2024pns, Li2023jnm, Xiang2024mtc}. 
Using \gls{md} simulations, Lin \textit{et al.} investigated delayed interstitial clustering in NiCoCrFe HEA, attributing this phenomenon to higher defect recombination efficiency and smaller interstitial loop binding energies compared to Ni~\cite{Gao2020acta}. 
In refractory \gls{bcc} \glspl{hea}, Chen \textit{et al.} developed semi-empirical interatomic potentials to study primary radiation damage, finding that \glspl{hea} exhibit more point defects but fewer dislocation loops compared to \gls{w}~\cite{Chen2023jnm, Qiu2023jpcm, Qiu2024jnm}.
In contrast, some researchers argue that \glspl{hea}' radiation resistance stems not from reduced primary damage via chemical disorder but rather from longer-time defect evolution~\cite{Deluigi2021acta}.
Thus, most perspectives on enhanced radiation resistance have focused on the energetics and diffusion behaviors of defects.
There remains a lack of understanding regarding how the morphology of displacement cascades affects defect production and evolution, particularly with high-energy \glspl{pka}.

This far, the majority of primary damage simulations have been conducted using \gls{eam} potentials. 
However, the limited accuracy of such traditional interatomic potentials with fixed functional forms for \gls{md} simulations often leads to challenges in simulating certain properties, such as the melting point, surface energies, and the energetics and structures of vacancy clusters, self-interstitial clusters, and dislocations~\cite{Byggmastar2020prm, Wang2022nf}.
Recently, \gls{ml} techniques have initiated a data-driven paradigm in developing interatomic potentials~\cite{behler2016jcp, Zhang2018prl, Deringer2019am, Mueller2020jcp, Noe2020arpc, Mishin2021am, Unke2021cr, batzner2022nc, Wang2024nc}.
The foundational framework of \glspl{mlp} is now well established, including general models built on extensive databases~\cite{takamoto2022cms, takamoto2022nc, chen2022ncs, deng2023nmi, zhang2024npj, Zhang2025npj}, sustainable approaches for constructing training sets \cite{Song2024nc}, sampling methods \cite{Zhang2020cpc, Fan2022jcp}, and related computational software \cite{Zhang2020cpc, wizard, Lindgren2024joss}.
These advances enable \glspl{mlp} to address challenges in simulating complex systems, particularly radiation damage~\cite{Wang2019apl, Byggmastar2019prb, Wang2022cms, Liu2023prb} in multi-component materials. 
For example, using the \gls{tabgap} model~\cite{Byggmastar2021prb, Byggmastar2020prm} for Mo-Nb-Ta-V-W, Wei \textit{et al.} investigated the effects of lattice and mass mismatch on primary radiation damage and high-dose defect accumulation~\cite{Wei2024jnm, Wei2024acta}.
Moreover, \glspl{mlp} have been employed to investigate segregation behaviors and defect dynamics relevant to irradiation~\cite{Byggmastar2021prb, jesper2024arxiv, He2024acta, Chen2024npj, jesper2024prm}.
While these studies provide important findings, they primarily focus on lower recoil energies and do not offer a comprehensive investigation of primary radiation damage across a broader spectrum of energies.

In addition to investigating the dynamic processes of radiation damage, quantifying the displacement damage caused by energetic particle interactions in materials is important in experimental and computer simulation studies.
The \gls{nrt} model~\cite{norgett_proposed_1975}, currently the international standard for quantifying radiation damage, has several limitations, such as overestimating the number of stable defects and underestimating the amount of atomic mixing caused by cascades~\cite{Nordlund2018jnm}.
The \gls{arc} model extends \gls{nrt} model to consider the major damage recombination and atomic mixing effects introduced by heat spikes~\cite{Nordlund2018nc}.
However, these models do not account for the effects of alloying elements’ concentrations, radii, and masses on collisions when predicting damage in alloy materials.

In this paper, we propose extended damage models that provide more physically realistic descriptions of primary radiation damage in multi-component materials.
Using the \gls{nep} approach~\cite{Fan2021prb, Fan2022jcp, Song2024nc}, we train a \gls{mlp} model for Mo-Nb-Ta-V-W system based on reference data from Byggm\"{a}star \textit{et al.}~\cite{Byggmastar2021prb, Byggmastar2020prm}. 
This \gls{nep} model achieves a computational speed of \num{1e7} atom-step/second with a single RTX 4090 GPU, which is comparable to the \gls{eam} potential \cite{Fan2022jcp, Song2024nc}. 
We evaluated the accuracy of this model and reported the results of a comprehensive investigation of \glspl{tde} in pure metals and MoNbTaVW \gls{hea}.
Additionally, we conduct a series of displacement cascade simulations in MoNbTaVW \gls{hea} and \gls{w} for the energy of \glspl{pka} ranging from 1 to 150 keV to compare the differences in the generation and evolution of point defects.
An extended \gls{arc} model was fitted to quantify the displacement damage in the MoNbTaVW \gls{hea}.
Furthermore, we performed 50 cascade simulations at the \gls{pka} energy of 150 keV to investigate the effect of different initial atom types, revealing that the subcascade splitting mechanism plays a crucial role in enhancing irradiation resistance.
With extended damage models and \gls{md} results, we explain how chemical composition affects primary radiation damage.
The present study of primary radiation damage in MoNbTaVW \gls{hea} offers a novel perspective on the irradiation-resistance mechanisms in \gls{bcc} alloys.

\section{Methodologies}  

\subsection{Extending damage models for multicomponent materials}
\gls{kp} established an early framework for \gls{dpa} by accounting for the number of displaced atoms as a function of energy~\cite{KP_1955}. 
The \gls{nrt} model~\cite{norgett_proposed_1975} improves upon the \gls{kp} formulation by replacing the original kinetic energy term with the damage energy to account for ionization effects, and introducing a displacement efficiency factor of 0.8.
The \gls{arc} model further extends the \gls{nrt} model by incorporating the effects of defect recombination through the introduction of an efficiency function~\cite{Nordlund2018nc}.
However, these models assume that all atoms are equivalent, neglecting variations in atomic concentrations, radii, and masses among different element types.

In this section, we propose an analytical extension of the \gls{kp} model for multicomponent systems.
By introducing the probability factor \(p_i\), which represents the probability that an atom of type \(i\) is selected as the target atom, we define an effective threshold displacement energy that accounts for the chemical complexity of multicomponent materials.
The derived expression is then generalized to both the \gls{nrt} and \gls{arc} models, enabling their application to chemically complex materials.
While this work applies the extended model to the MoNbTaVW \gls{hea} as a representative case, its analytical foundation ensures applicability to a broad range of multicomponent materials, such as ferritic/martensitic steels and medium-entropy alloys.
This model is represented by the following equation:
\begin{equation}
    \nu(E) = \nu(E - T) + \nu(T).
\end{equation}
Here, \(E\) represents the initial kinetic energy, and \(T\) denotes the energy transferred during a collision. \(\nu(E)\) represents the number of displaced atoms corresponding to energy \(E\). Considering that different types of atoms have distinct \(E_\mathrm{d}\) (\glspl{tde}), we introduce \(\omega_{ij}(E)\) to describe the number of displaced atoms when atoms of types \(i\) and \(j\) collide:
\begin{equation}
    \omega_{ij}(E) = \nu_i(E-T) + \nu_j(T).
\end{equation}
\(\nu_i(E)\) represents the number of displaced atoms produced by an atom of type \(i\) with energy \(E\). For a collision between two atoms of types \(i\) and \(j\), the maximum transferable energy is given by \( T_{max} = \Lambda_{ij} E \).  The parameter \(\Lambda_{ij}\) is defined as:
\begin{equation}
    \Lambda_{ij} = \dfrac{4 M_i M_j}{(M_i + M_j)^2},
\end{equation}
where \(M_i\) and \(M_j\) represent the masses of atoms \(i\) and \(j\) respectively. The average number of displaced atoms denoted by \( \overline{\omega}_{ij}(E) \):
\begin{equation}
    \overline{\omega}_{ij}(E) = \int_0^{\Lambda_{ij} E} \left[ \overline{\nu}_i(E - T) + \overline{\nu}_j(T) \right] \frac{K_{ij}(E,\ T)}{\sigma_{ij}(E)} \mathrm{d}T.
\end{equation}
Here, \( K_{ij}(E,\ T)dT\) represents the differential cross-section for energy transfer, while \( \sigma_{ij}(E) \) is the total scattering cross-section. Based on the fundamental assumptions of the \gls{kp} model, the energy transfer cross-section is determined by the hard sphere model. For hard sphere collisions,
\begin{equation}
    \frac{K_{ij}(E,\ T)}{\sigma_{ij}(E)} = \frac{1}{\Lambda_{ij} E}.
\end{equation}
Thus simplifying the expression for the average number of displaced atoms, \( \overline{\omega}_{ij}(E) \), to:
\begin{equation}
    \overline{\omega}_{ij}(E) = \frac{1}{\Lambda_{ij} E} \int_0^{\Lambda_{ij} E} \left[ \overline{\nu}_i(E - T) + \overline{\nu}_j(T) \right] \, \mathrm{d}T.
\end{equation}
Similarly,
\begin{equation}
    \overline{\omega}_{ji}(E) = \frac{1}{\Lambda_{ij} E} \int_0^{\Lambda_{ij} E} \left[ \overline{\nu}_j(E - T) + \overline{\nu}_i(T) \right] \, \mathrm{d}T.
\end{equation}
Hence,
\begin{align}
   \overline{\nu}(E) &= \sum_{i}^{n} p_i \sum_{j}^n \frac{p_j}{\Lambda_{ij} E} \int_0^{\Lambda_{ij} E} \left[ \overline{\nu}_i(E - T) + \overline{\nu}_j(T) \right] \, \mathrm{d}T \nonumber \\
   & = \frac{1}{2} \left[ \sum_{i}^{n} \sum_{j}^n \frac{p_i p_j}{\Lambda_{ij} E} \int_0^{\Lambda_{ij} E} \left[ \overline{\nu}_i(E - T) + \overline{\nu}_j(T) \right] \, \mathrm{d}T \right. \nonumber \\
   &\quad + \left. \sum_{j}^{n} \sum_{i}^n \frac{p_j p_i}{\Lambda_{ij} E} \int_0^{\Lambda_{ij} E} \left[ \overline{\nu}_j(E - T) + \overline{\nu}_i(T) \right] \, \mathrm{d}T \right] \nonumber \\
   & = \frac{1}{2} \left[ \sum_{i}^{n} \sum_{j}^n \frac{p_i p_j}{\Lambda_{ij} E} \int_0^{\Lambda_{ij} E} \left[ \overline{\nu}_i(E - T) + \overline{\nu}_i(T) \right] \, \mathrm{d}T \right. \nonumber \\
   &\quad + \left. \sum_{j}^{n} \sum_{i}^n \frac{p_j p_i}{\Lambda_{ij} E} \int_0^{\Lambda_{ij} E} \left[ \overline{\nu}_j(E - T) + \overline{\nu}_j(T) \right] \, \mathrm{d}T \right] \nonumber \\
   & = \sum_{i}^{n} \sum_{j}^n \frac{p_i p_j}{\Lambda_{ij} E} \int_0^{\Lambda_{ij} E} \left[ \overline{\nu}_i(E - T) + \overline{\nu}_i(T) \right] \, \mathrm{d}T.
\label{eq8}
\end{align}

The probability \(p_i\) of an atom of type $i$ being the target atom during a binary collision is expressed as:
\begin{equation}
    p_{i} = \dfrac{c_{i} r_{i}^2}{\sum_{j}^{n} c_{j} r_{j}^2},
\end{equation}
where \(c_i\) is the atomic concentration, and \(r_i\) is the atomic radius. The \gls{kp} model assumes a random and uniform atomic distribution, disregarding the influence of atomic arrangement within the solid. 
Under this assumption, atomic interactions are simplified to two-dimensional plane collisions rather than three-dimensional interactions.
The term \(c_i r_i^2\) represents the combined effects of atomic abundance and size for type \(i\), while the denominator normalizes the probabilities across all atom types. 
Ordered structures could potentially impact the results and deserve detailed investigation in future studies. 

Through a change of integration variables applied to Eq.~\ref{eq8}, we can derive the following expression:
\begin{equation}
    \overline{\nu}(E) = \sum_{i, j} \frac{p_i p_j}{\Lambda_{ij} E} \left[ \int_{(1-\Lambda_{ij})E}^{E} \overline{\nu}_i(T) \, \mathrm{d}T + \int_0^{\Lambda_{ij} E} \overline{\nu}_i(T) \, \mathrm{d}T \right].
\label{d_e}
\end{equation}
Clearly, when the system consists of a single atom type, Eq.~\ref{d_e} simplifies to:
\begin{equation}
    \overline{\nu}(E) = \frac{2}{E} \int_0^{E} \overline{\nu}(T) \, \mathrm{d}T.
\end{equation}
Solving this equation yields,
\begin{align}
   \overline{\nu}(E) = CE,
\end{align}
where \(C\) represents a constant. Consequently, we hypothesized that \(\overline{\nu}(E)\) has a linear relationship with respect to \(E\) in Eq.~\ref{d_e}, and this hypothesis can be verified by substituting into the equation. Taken together, we propose a modified defect production model:
\begin{equation}
\overline{\nu}(E) =
\left[
\begin{array}{rl}
0\hspace{4pt}, & \quad E < E_\mathrm{d} \\[2mm]
1\hspace{4pt}, & E_\mathrm{d} < E < E_\mathrm{multi} \\[2mm]
\frac{E}{E_\mathrm{multi}}, & E_\mathrm{multi} < E < \infty
\end{array}.
\right]
\end{equation}

In light of the above derivation, we define the effective threshold energy for multi-component systems as follows:
\begin{equation}
\label{em}
    E_\mathrm{multi} = \sum_{i}^{n}p_{i}(E_\mathrm{d}^{i} + \sum_{j}^{n}p_{j}E_\mathrm{d}^{j}/\Lambda_{ij})
\end{equation}
\(E_\mathrm{d}^{i}\) represents the \gls{tde} for an atom of type \(i\) in multi-component materials. Similarly, we can derive the \gls{nrt} model~\cite{norgett_proposed_1975}:
\begin{equation}
N_{\text{d, NRT-dpa}}(T_\mathrm{d}) = 
\left[
\begin{array}{rl} 
0\hspace{8pt},& \quad T_\mathrm{d} < E_\mathrm{d}                      \\[2mm]
1\hspace{8pt},& E_\mathrm{d} < T_\mathrm{d} < \frac{E_\mathrm{multi}}{0.8}   \\[2mm]
\frac{0.8T_\mathrm{d}}{E_\mathrm{multi}},  & \frac{E_\mathrm{multi}}{0.8} < T_\mathrm{d} < \infty
\end{array}
\right]
\end{equation}
Here, $T_d$ represents the damage energy, which is the kinetic energy available for creating atomic displacements. The damage energy for a single ion is determined by subtracting the energy lost to electronic interactions, such as ionization, from the total ion energy. The \gls{arc} model~\cite{Nordlund2018nc} can be extended as:
\begin{equation}
N_{\text{d, arc-dpa}}(T_\mathrm{d}) = 
\left[
\begin{array}{rl} 
0\hspace{30pt},& \quad T_\mathrm{d} < E_\mathrm{d}                      \\[2mm]
1\hspace{30pt},& E_\mathrm{d} < T_\mathrm{d} < \frac{E_\mathrm{multi}}{0.8}   \\[2mm]
\frac{0.8T_\mathrm{d}}{E_\mathrm{multi}}\xi_{\text{arcdpa}}(T_\mathrm{d}),  & \frac{E_\mathrm{multi}}{0.8} < T_\mathrm{d} < \infty
\end{array}
\right]
\end{equation}
with the function \( \xi_{\text{arc-dpa}}(T_\mathrm{d}) \) given by:
\begin{equation}
\xi_{\text{arcdpa}}(T_\mathrm{d}) = \frac{1 - c_{\text{arcdpa}}}{(E_\mathrm{multi}/0.8)^{b_{\text{arcdpa}}}}T_\mathrm{d}^{b_{\text{arcdpa}}} + c_{\text{arcdpa}}.
\end{equation}
The parameters \(b_\text{arcdpa}\) and \(c_\text{arcdpa}\) are material constants that can be determined for a given material from \gls{md} simulations or experiments.

\subsection{Machine-learned potential}

\begin{figure}[h] 
\centering 
\includegraphics[width=\columnwidth]{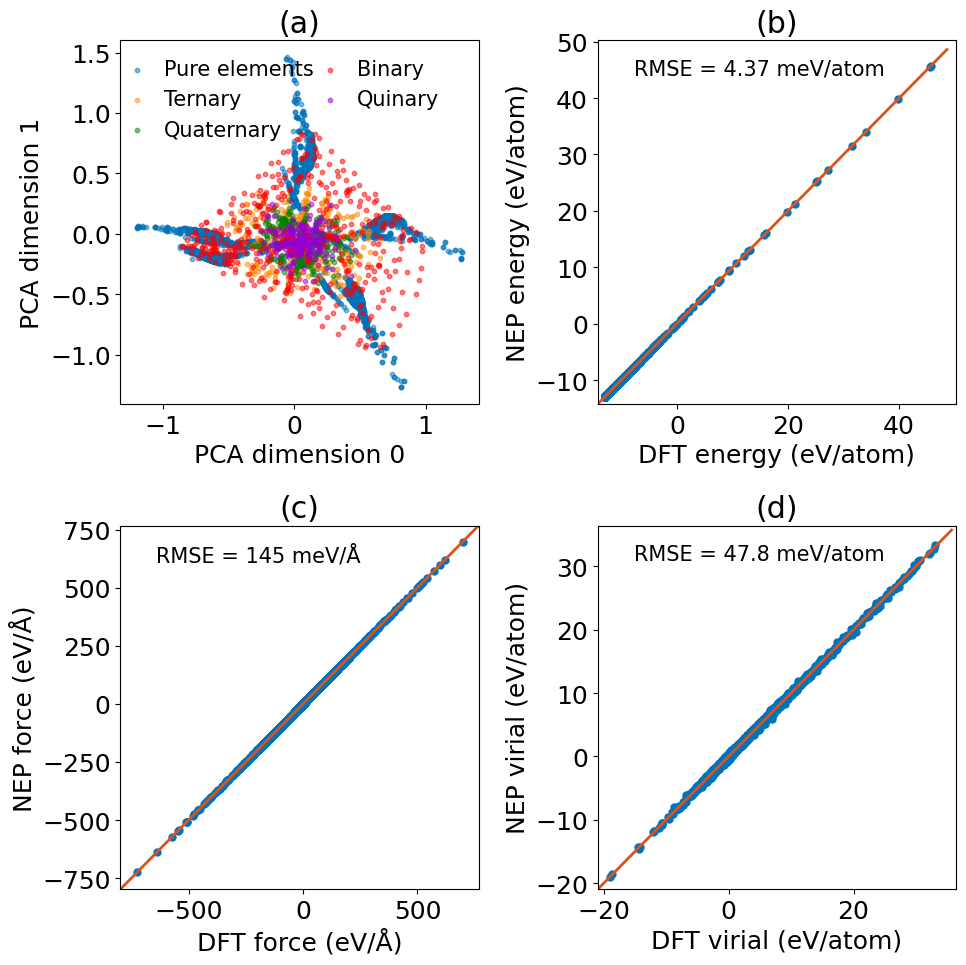}
\caption{(a) Distribution of the training dataset in the two-dimensional (2D) principal component (PC) space of the descriptor. (b) Energy, (c) force, and (d) virial as calculated from the \gls{nep} model compared with the training data.} 
\label{fig:nep}
\end{figure}

We utilized the alloy training data set from Ref.~\cite{Byggmastar2021prb} and all pure metal structures from Ref.~\cite{Byggmastar2020prm}.
Figure~\ref{fig:nep}(a) shows the distribution of the training dataset in the \gls{2d} \gls{pc} space of the descriptor.
For pure metals, the dataset includes elastically distorted unit cells of \gls{bcc}, high-temperature \gls{bcc} crystals, structures containing vacancies and self-interstitial atoms, as well as surfaces and liquids.
The dataset also includes a sampling of all alloy chemical compositions from binary to five-element random alloys to ensure transferability to arbitrary compositions. Various ordered alloys from binary to quinary compositions are also included. For defects in alloys, the dataset primarily focuses on the MoNbTaVW \gls{hea}  with up to five vacancies or self-interstitial atoms. 
To capture short-range repulsion, the dataset also includes \gls{hea} crystals with a randomly inserted interstitial atom positioned near—but not too close to—a neighboring atom.
Liquid structures span all equiatomic binary to quinary combinations. For more details, see the Supplementary document of Ref.~\cite{Byggmastar2021prb}.
These structures, totaling 21 672 structures and 334 858 atoms, ensure that the \gls{mlp} can accurately simulate the elastic, thermal, and defect properties, as well as surface energetics, the melting process, and the structure of the liquid phase.
For detailed information on the strategy for constructing the training structures and the \gls{dft} calculations, please refer to Refs.~\cite{Byggmastar2020prm, Byggmastar2021prb}.

The \gls{nep} model \cite{Fan2021prb, Fan2022jcp} with the NEP4 flavor \cite{Song2024nc} for Mo-Nb-Ta-V-W \gls{hea} was trained using the \textsc{gpumd} package \cite{Fan2017cpc}. 
Figure~\ref{fig:nep}(b)-(d) compare the predicted energy, force, and virial values by the \gls{nep} model and those from quantum-mechanical \gls{dft} calculations for the training set.
The \glspl{rmse} of energy, force, and virial for the \gls{nep} model are 4.37 meV/atom, 145 meV/\AA, and 47.8 meV/atom, respectively. 
Moreover, with a single RTX 4090 GPU, this \gls{nep} model achieves a computational speed of \num{1e7} atom-step/second and can simulate about 8 million atoms, which is sufficient for the computational efficiency required for cascade simulations.

The major hyperparameters in the NEP4 model are chosen as follows. The cutoff radii for both the radial and angular descriptor components are 5 \AA. 
For both the radial and angular descriptor components, there are 7 radial functions, each being a linear combination of 9
Chebyshev-polynomial-based basis functions with trainable expansion coefficients.
The loss function is a weighted sum of the \glspl{rmse} of energy, force, and virial with relative weights of 1, 1, and 0.1, respectively. 
The number of neurons in the hidden layer of the neural network is $80$. 
All other parameters are set to the default values. 

For accurate characterization of short-range repulsive forces, we combine the \gls{zbl} potential with the \gls{nep} model following the \gls{nep}-\gls{zbl} scheme~\cite{Liu2023prb}.
The total site energy $U_i$ on atom $i$ is then
\begin{equation}
U_{i} = U_{i}^{\rm NEP} \left( \{ q_{\nu}^{i} \} \right) 
+ \frac{1}{2} \sum_{j\neq i} U_{\rm ZBL}(r_{ij}).
\end{equation}
The repulsive ZBL potential is a screened Coulomb potential
\begin{equation}
U_{\rm ZBL}(r_{ij})=\frac{1}{4\pi\epsilon_{0}}\frac{Z_iZ_je^2}{r_{ij}}\phi(r_{ij}/a)f_{\rm c}(r_{ij}),
\end{equation}
where
\begin{equation}
a = \frac{0.46848}{Z_i^{0.23} + Z_j^{0.23}}.
\end{equation}
Here, $\epsilon_0$ is the vacuum dielectric constant, $Z_ie$ is the nuclear charge of atom $i$, and $r_{ij}$ is the distance between atoms $i$ and $j$. For the cutoff function, $f_{\rm c}(r_{ij})$, we take it as the Tersoff one \cite{Tersoff1989prb} with an inner cutoff of 1.0 \AA{} and an outer cutoff of 2.0 \AA. The screening function $\phi$ uses parameters from Ref.~\cite{Byggmastar2021prb}, which were specifically optimized for all element pairs in MoNbTaVW.
Beyond 2 \AA, there is only \gls{nep} in action.
Below 2 \AA, \gls{nep} is trained against the difference between DFT and ZBL. The training dataset contains a lot atom pairs in the transition region (1 to 2 \AA), ensuring smooth dimer curves as shown in the Supplementary Material Figs. 2 and 3.
Static calculations were performed using \textsc{gpumd-wizard}~\cite{wizard}, \textsc{ase}~\cite{Larsen2017jpcm}, and \textsc{calorine}~\cite{Lindgren2024joss}. 
The training and validation results for this model are publicly accessible at the Zenodo repository \cite{wizard}.

\subsection{Molecular dynamic simulations}

The \gls{md} simulations of collision cascades were performed using the GPUMD package \cite{Fan2017cpc}. 
The MoNbTaVW \gls{hea} simulation cell was constructed by creating an equimolar, random mixture of Mo, Nb, Ta, V, and W elements within a defined \gls{bcc} crystal structure. 
To prepare the system for initiating a cascade, we equilibrate it under the isothermal-isobaric ensemble for 30 ps, with a target temperature of 300 K and a target pressure of 0 GPa. 
All three directions are treated as periodic. 
High-energy particles are created at the center of the simulation box. 
Tungsten, a candidate material for fusion reactors, was used as a reference for comparison with the MoNbTaVW \gls{hea}. 
The \gls{pka} energies, numbers of simulation steps, box lengths and numbers of atoms, are presented in Table~\ref{table:simulation_para}. 
For tungsten, each simulation was run 10 times. 
To achieve statistical convergence, \glspl{hea} simulations were run at least 20 times for each energy.
At $E_{\rm PKA} = 30,~40,~50,~150~\mathrm{keV}$, 50 simulations were conducted to investigate the impact of different \gls{pka} types with 10 independent simulations for each PKA type (Mo, Nb, Ta, V, W).
All reported results are statistical averages over these independent simulations.
The initial momenta of high-energy particles are chosen in the high-index direction $\langle135\rangle$ to avoid channeling effects~\cite{Stoller2000jnm, Gao2010prb, Stoller2012, Ortiz2018cms, Gao2021nc, Fu2019jnm}, ensuring that the results are not influenced by crystallographic directions.
It is acknowledged that the incident angle, such as channeling or near-channeling directions, may affect defect formation. 
Further research and comprehensive investigations are therefore important.
Atoms within a thickness of $3a_0$ of the boundaries of the simulation boxes are maintained at 300 K using the Nose-Hoover chain thermostat \cite{Martyna1992jcp}. 
The integration time step is dynamically determined so that the fastest atom can move at most 0.015 \AA~  within one step, with an upper limit of 1 fs also set. 
Electronic stopping \cite{Kai1995cms} was applied as a frictional force on atoms with a kinetic energy over 10 eV, using data from the SRIM-2013 code \cite{ziegler2010srim, ziegler2013}. 

\begin{table}[htbp]
  \centering \setlength{\tabcolsep}{1.6mm} 
  \caption{Simulation parameters for W and Mo-Nb-Ta-V-W systems: the \gls{pka} energy $E_{\rm PKA}$ in units of keV, the damage energy $T_{\rm d}$ in units of keV, the simulation time in units of ps, the number of bcc unit cells $L$ in the simulation box, and the number of atoms $N$.}
  \begin{tabular}{lllllll}
    \hline
    \hline
     &$E_{\rm PKA}$  & $T_{\rm PKA}$ & Time & $L$ & $N$ (Million) & \\
    \hline
     MoNbTaVW  & 1   &  0.84  & 50   & 100   & 2   \\
               & 5   &  4.2   & 50   & 100   & 2   \\
               & 10  &  8.4   & 50   & 100   & 2   \\
               & 20  &  16    & 50   & 150   & 6.75  \\
               & 30  &  25    & 50   & 150   & 6.75  \\
               & 40  &  32    & 50   & 150   & 6.75  \\
               & 50  &  40    & 50   & 150   & 6.75  \\
               & 100 &  78    & 100  & 200   & 16  \\
               & 150 &  115   & 100  & 240   & 27.648 \\ 
    W          & 1   &  0.77  & 50   & 50    & 0.25    \\
               & 5   &  3.9   & 50   & 50    & 0.25    \\
               & 10  &  7.7   & 50   & 50    & 0.25    \\
               & 20  &  15    & 50   & 100   & 2   \\
               & 30  &  23    & 50   & 100   & 2   \\
               & 40  &  30    & 50   & 100   & 2   \\
               & 50  &  37    & 50   & 100   & 2   \\
               & 100 &  75    & 100  & 150   & 6.75  \\
               & 150 &  112   & 100  & 200   & 16 \\
               
     \hline
     \hline
    \end{tabular}
      \label{table:simulation_para}
\end{table}

All \gls{tde} ($E_\mathrm{d}$) calculations were performed with the \gls{nep} model at 300 K. 
The simulation box was a $12 \times 13 \times 14$ supercell containing \num{4368} atoms.  
We sampled 500 random recoil directions in pure metals and 1000 random recoil directions per element in the \gls{hea} to obtain a converged average ($E_\mathrm{d}$) at 300 K. 
The simulation methods are similar to the cascade simulations described above, with adaptive time step and cooling down by one lattice atomic layer at boundaries.
The atom in the center of the system is selected as the \gls{pka} and given an initial velocity in a random direction.
Displacement simulations (6 ps) with increasing recoil energies (increment 2 eV) were performed until a stable \gls{fp} is formed. 
'FP' refer to pair of interstitial and vacancy identified using the Wigner-Seitz cell method.
Then, the $E_\mathrm{d}$ of \gls{pka} is decreased by 1 eV to determine the final $E_\mathrm{d}$.
Before every new random crystal direction was sampled, we randomly shifted the simulation system to obtain a new chemical environment for the recoil event. 
Due to the low energy, no electronic stopping was used in these simulations. 

We used the \textsc{ovito} package \cite{ovito} for defect analysis and visualization. 
Interstitials and vacancies were identified using the Wigner-Seitz cell method. Defects were clustered using a cutoff radius set between the second- and third-nearest neighbors for vacancies, and between the third and fourth for interstitials, with clusters containing at least two defects.
Furthermore, the sizes of the defect clusters in this study were determined based on the net defect count, which results from the difference between the numbers of interstitials and vacancies. 

\section{Results}

\subsection{Validating the machine-learned potential}

\begin{table*}[t] 
\centering \setlength{\tabcolsep}{5.8mm} 
\caption{Basic properties of the pure bcc metals and MoNbTaVW: Energy per atom of the bcc phase~$E_\mathrm{bcc}~\rm{(eV/atom)}$, the mixing energy per atom ~$E_\mathrm{mix}~\rm{(eV/atom)}$, lattice constant~$a~\rm{(\AA)}$, bulk modulus~$B~\rm{(GPa)}$, melting temperature~$T_\mathrm{melt}~\rm{(K)}$, and average threshold displacement energies $E_\mathrm{d}~\rm{(eV)}$ at 300 K. All properties are computed with the \gls{nep} model and compared with experiments or DFT from the literature (italic font).}
\begin{tabular}{lllllll} 
\hline 
\hline 
&V & Nb & Mo & Ta & W & HEA \\
\hline 
$E_\mathrm{bcc}$ & -8.992 & -10.217 & -10.937 & -11.812 & -12.956 & -11.026 \\
&\textit{-8.992}\cite{Byggmastar2020prm} & \textit{-10.216}\cite{Byggmastar2020prm} &\textit{-10.936}\cite{Byggmastar2020prm} &\textit{-11.812}\cite{Byggmastar2020prm} &\textit{-12.957}\cite{Byggmastar2020prm} & \\
$E_\mathrm{mix}$ & 0 & 0 & 0 & 0 & 0 & -0.043\\
$a$ & 2.997 & 3.308 & 3.164 & 3.320 & 3.186 & 3.195 \\
&\textit{2.997}\cite{Byggmastar2020prm} & \textit{3.307}\cite{Byggmastar2020prm} &\textit{3.163}\cite{Byggmastar2020prm} &\textit{3.319}\cite{Byggmastar2020prm} &\textit{3.185}\cite{Byggmastar2020prm} & \\
&\textit{3.024}\cite{Rumble2019crc} & 
\textit{3.300}\cite{Rumble2019crc} & 
\textit{3.147}\cite{Rumble2019crc} & 
\textit{3.303}\cite{Rumble2019crc} & 
\textit{3.165}\cite{Rumble2019crc} & 
\textit{3.183}~\cite{Senkov2010intermetallics} \\
$B$ & 195 & 168 & 261 & 192 & 309 & 219\\
&\textit{187}\cite{Byggmastar2020prm} & \textit{171}\cite{Byggmastar2020prm} &\textit{259}\cite{Byggmastar2020prm} &\textit{196}\cite{Byggmastar2020prm} &\textit{303}\cite{Byggmastar2020prm} & \textit{218}\cite{Byggmastar2021prb}\\
$T_\mathrm{melt}$ &2170&2580&2770&3000&3570&2840\\
&\textit{2183}\cite{Rumble2019crc} & 
\textit{2750}\cite{Rumble2019crc} & 
\textit{2895}\cite{Rumble2019crc} & 
\textit{3290}\cite{Rumble2019crc} & 
\textit{3687}\cite{Rumble2019crc} &  \\
$E_\mathrm{d}$ & 54 & 70 & 99 & 77 & 109 & 51\\
\hline 
\hline 
\end{tabular}
\label{table:property}
\end{table*}

Table~\ref{table:property} lists basic properties of pure metals and the \gls{hea} as calculated by the \gls{nep} model, and compares these to experimental and \gls{dft} data from literature.
The \gls{nep} model shows a satisfactory agreement in the predictions of energies, lattice parameters, and bulk moduli.
For the \gls{hea}, these properties are determined as averages from 50 relaxed systems, each containing 2000 atoms.
The results indicate that W and Mo are elastically the stiffest, and alloying them with softer metals leads to a corresponding reduction in elastic stiffness.
Furthermore, we compare predictions from \gls{dft} and the \gls{nep} model for equiatomic random alloys, including energy–volume relations, bulk moduli, and mixing energies at equilibrium volumes, as shown in Supplementary Figs.~1(a) and 1(b), demonstrating the model’s accuracy in capturing phase stability and elastic response across binary to quinary compositions.
The melting point was calculated using the solid-liquid coexistence method~\cite{luo2004jcp}. 
Bi-phase systems containing 13,500 atoms, with half of the atoms in the liquid phase and the other half in the solid bcc phase, were simulated at temperatures near the melting point, with pressures maintained at 0 GPa.
It should be noted that our simulations provide only a single point on the phase diagram between the solidus and liquidus for the \gls{hea}. 

\begin{figure*}[htbp]
\centering
\includegraphics[width=1.5\columnwidth]{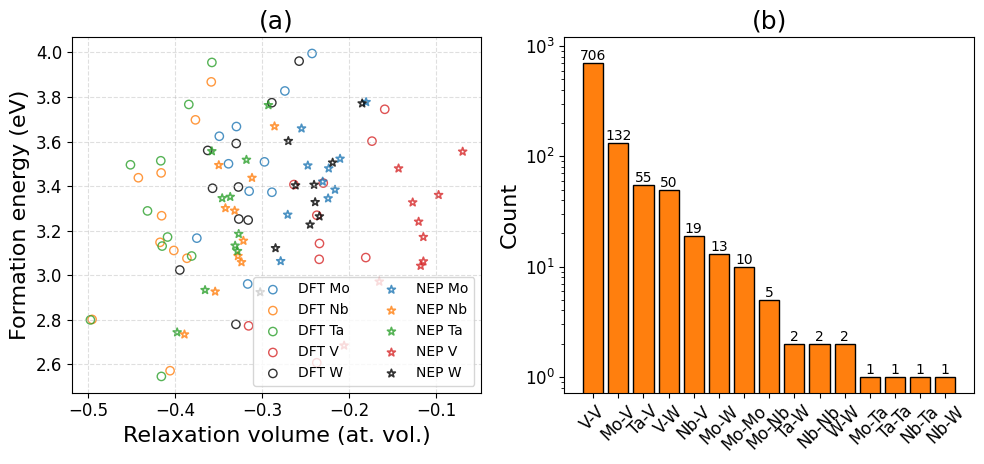}
\caption{(a) The single-vacancy data are separated by element. (b) Distribution of stable self-interstitial dumbbell configurations in
MoNbTaVW obtained with the \gls{nep}.}
\label{fig:point-defct}
\end{figure*}

The point defect properties for pure metals~\cite{Ma2019prm1, Ma2019prm2, Ma2019prm3} are presented in the Supplementary Materials, with computational details described in Ref.~\cite{Liu2023prb}.
For the \gls{hea}, the chemical complexity introduces numerous possible configurations for these simple defects, requiring a statistical treatment~\cite{Byggmastar2021prb}.
Here, we focus on randomly ordered \glspl{hea} with randomly added single vacancies and interstitials, followed by relaxation.
The vacancies are created in ten different \gls{hea} systems, using the same systems as in the DFT calculations in Ref.~\cite{Byggmastar2021prb} to allow a direct comparison. 
For each vacancy system, reference bulk systems are created by filling the vacancy with each element separately, resulting in data for 50 different vacancies.
Figure~\ref{fig:point-defct}(a) shows the formation energies of single vacancies in the \gls{hea}, comparing results from \gls{dft} calculations and the \gls{nep} model. 
The average vacancy formation energy is 3.33 eV according to \gls{dft} and 3.29 eV according to the \gls{nep}. 
For self-insterstitials, we relaxed 1000 HEA systems each containing one randomly inserted interstitial atom.
The distributions of the relaxed dumbbell configurations are presented in Figure~\ref{fig:point-defct}(b).
The observed dominance of V-containing interstitial dumbbells, especially pure V-V dumbbells, is consistent with prior MD and DFT studies~\cite{Zhao2020jmst, Byggmastar2021prb}. 
Overall, our model demonstrates reliable performance in describing point defects, consistently aligning with observed phenomena and \gls{dft} results.

\begin{figure*}[htbp]
\centering
\includegraphics[width=1.8\columnwidth]{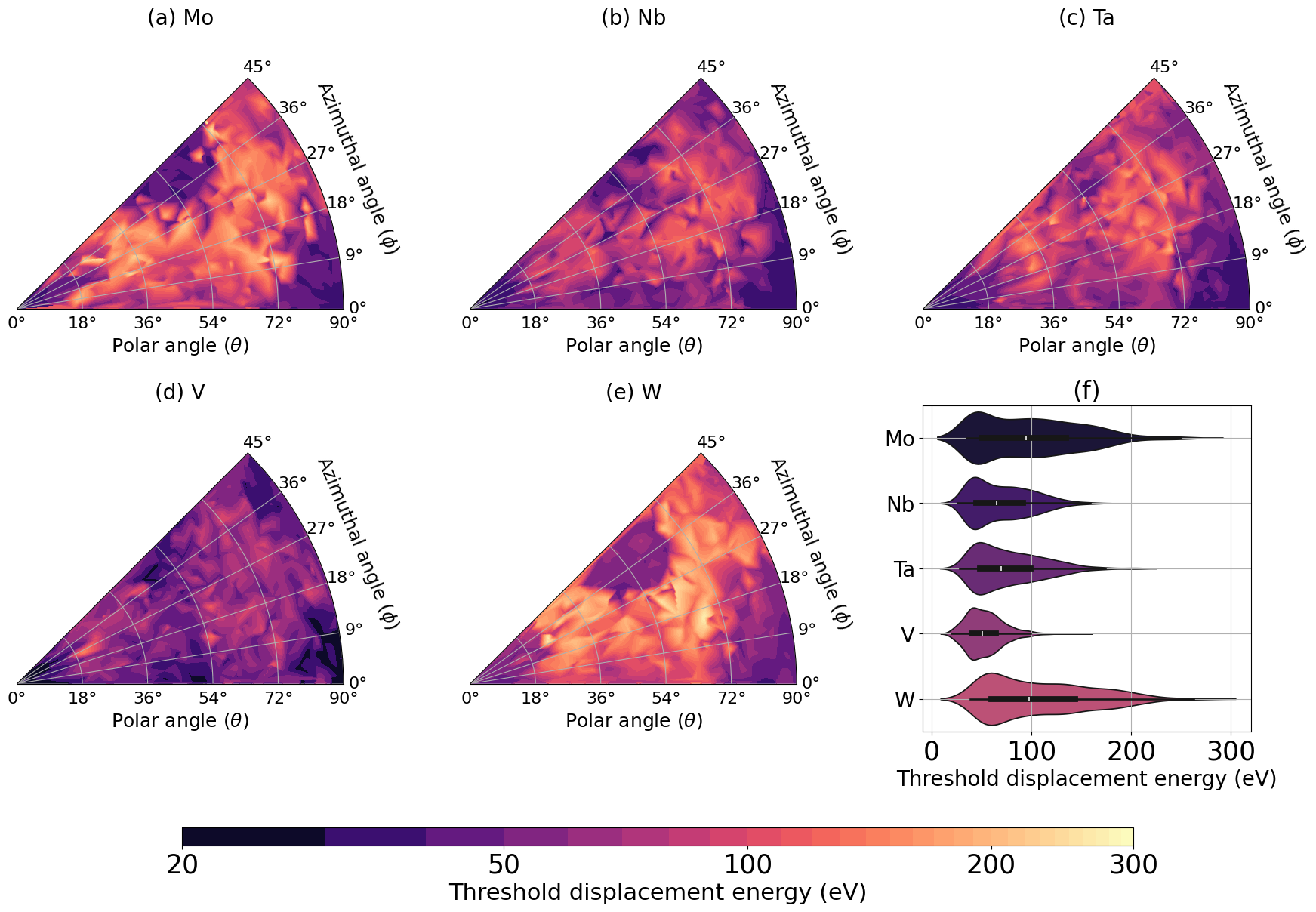}
\caption{Threshold displacement energies (TDEs) maps for pure metals: (a) Mo, (b) Nb, (c) Ta, (d) V, (e) W, and (f) the distributions of TDEs. Each map includes all calculated directions and is interpolated using the linear interpolation method.}
\label{fig:tde-metal}
\end{figure*}

\begin{figure*}
\centering
\includegraphics[width=1.8\columnwidth]{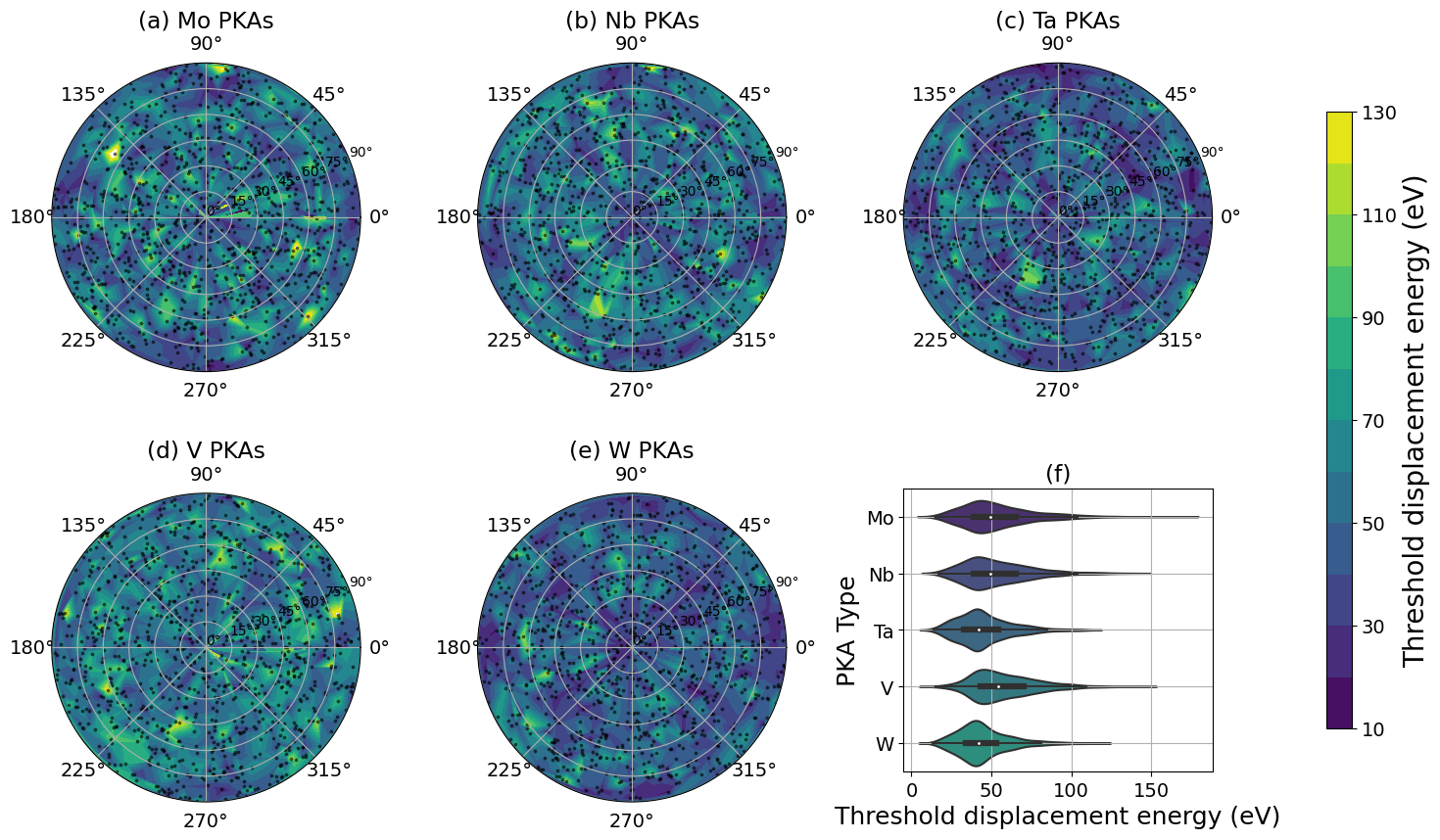}
\caption{Threshold displacement energies (TDEs) maps for MoNbTaVW high-entropy alloy, showing the response of different atomic PKAs: (a) Mo, (b) Nb, (c) Ta, (d) V, (e) W, and (f) the distributions of $E_\mathrm{d}$ for the five PKA types in MoNbTaVW. Each map includes all calculated directions and is interpolated between the black dots using the linear interpolation method.}
\label{fig:tde}
\end{figure*}

\subsection{Frenkel pair generation and evolution}

The \gls{tde} is the most fundamental property of radiation damage, representing the minimum recoiling kinetic energy required to displace an atom to create one or more stable defects.
In this study, we simulate \glspl{tde} for MoNbTaVW \gls{hea} and pure metals, with the average $E_\mathrm{d}$ presented in Table~\ref{table:property}. 
The \glspl{tde} obtained by the \gls{nep} model in the $\langle100\rangle$ direction are consistent with experimental data~\cite{Jung1975re, Maury1979prb, Maury1978re}, as shown in the Supplementary Materials.
Figure~\ref{fig:tde-metal} shows the angular maps of \glspl{tde} for pure metals.
Among pure metals, W has the highest \gls{tde}, followed by Mo, while V atoms are the easiest to displace from perfect lattice positions.
This trend is consistent with the formation energies of interstitials in pure metals.
Moreover, the \glspl{tde} of W and Mo are distributed across a broader range of values and exhibit greater dependence on lattice directions, as shown in Figure~\ref{fig:tde-metal}(f). 
Figure~\ref{fig:tde} illustrates the \gls{tde} maps of the five \gls{pka} type in MoNbTaVW \gls{hea}.
In contrast, the \gls{tde} distribution for the MoNbTaVW \gls{hea} is closer to a Gaussian shape and shows less angular dependence.
This is partly due to lattice distortions in \glspl{hea}, which make the local atomic environment more uniform.
The \gls{tde} values for the five \gls{pka} types are: V 58, Nb 53, Mo 54, Ta 45, and W 45, indicating an inverse relationship between \gls{tde} and atomic mass, where lighter \glspl{pka} require higher energies for displacement.
In a total of 5000 recoil directions sampled, the probability of each type of atom forming stable defects was as follows: V, 85\%; Nb, 3\%; Mo, 9\%; Ta, 0.6\%; W, 2.4\%.
This is consistent with Ref.~\cite{Byggmastar2021prb}, which reports the trend of V-containing dumbbells having the lowest formation energies and attributes this to V being the smallest atom, thereby preferring shorter interatomic bonds compared to other elements.

\begin{figure}[h]
\centering
\includegraphics[width=\columnwidth]{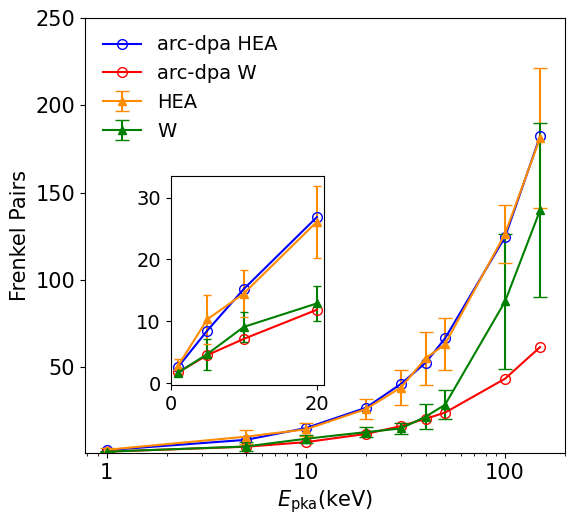}
\caption{Average number of residual point defects and the corresponding arc-dpa model results. Error bars are given as the standard deviations. The inset in the figure shows a duplicate view of the low-value data for better visibility and comparison.}
\label{fig:fps}
\end{figure}

A series of atomic collision cascade simulations were run over a \gls{pka} energy range of 1 keV to 150 keV.
Figure~\ref{fig:fps} shows the average number of \glspl{fp} that survived at the final stage of the cascade simulation in both the MoNbTaVW \glspl{hea} and pure W.
We use the extended \gls{arc} model to quantify the amount of displacement damage. 
The damage energies ($T_{\rm d}$), defined as the total ion energy minus the energy lost to electronic interactions, are listed in Table~\ref{table:simulation_para}.
Notably, pure tungsten loses more energies to electronic interactions during irradiation compared to the \gls{hea}, which influences the formation and survival of defects.
For pure W, the fitted \gls{arc} parameters $b_\text{arcdpa} = -0.56$ and $c_\text{arcdpa} = 0.12$, as reported in Ref.~\cite{Nordlund2018nc}, are used for comparison.
In lower energy regions, the average number of surviving \glspl{fp} obtained from our simulations agrees well with the predictions of the \gls{arc} model, whereas in higher energy regions, it significantly exceeds the model predictions.
This trend is consistent with the findings reported in Ref.~\cite{Nordlund2018nc}.
The \gls{hea} exhibits a higher number of surviving \glspl{fp} than pure W across the entire considered energy range.
From the \glspl{tde} for the five \gls{pka} types, the effective threshold energy for multi-component systems, $E_\mathrm{multi} = 108$ eV, is calculated using Eq.~\ref{em}, indicating that atoms in MoNbTaVW are harder to displace than predicted by the initial models.
By fitting the \gls{md} results, we determine the parameters for the MoNbTaVW \gls{hea} as $b_\text{arcdpa} = -0.73$ and $c_\text{arcdpa} = 0.21$. 
In contrast, while the original model also yields $b_\text{arcdpa} = -0.73$, it yields a lower $c_\text{arcdpa}$ of 0.19, indicating an underestimation of damage production at high displacement energies, as $\xi_{\text{arcdpa}} \to c_\text{arcdpa}$ when $T_\mathrm{d} \to \infty$.
The extended damage model provides more physically realistic parameters and emphasizes the impact of chemical complexity on damage production.
Compared to pure W, the \gls{md} results and the fitted model for HEAs show excellent consistency, without noticeable energy-dependent variations, which in pure metals are typically associated with the direct formation of large interstitial clusters~\cite{Setyawan2015jnm}.

\begin{figure*}[htb]
\centering
\includegraphics[width=1.5\columnwidth]{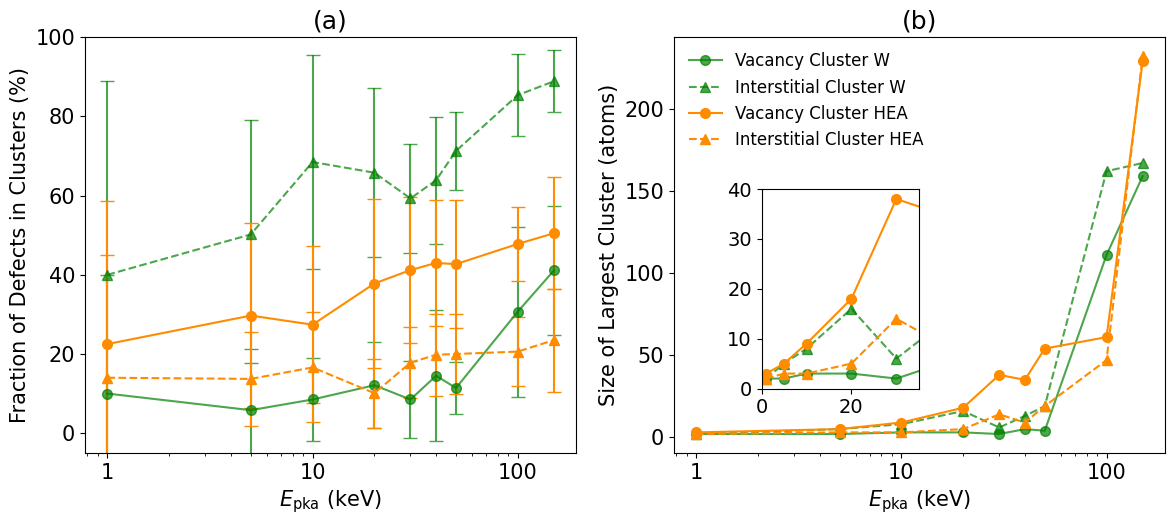}
\caption{(a) Average clustered fraction of vacancies and interstitials. Error bars represent standard deviations. (b) Number of atoms in the largest vacancy and interstitial clusters. The inset in the figure shows a duplicate view of the low-value data for better visibility and comparison.}
\label{fig:cluster-all}
\end{figure*}

Figure~\ref{fig:cluster-all}(a) shows the clustered fraction of surviving vacancies and interstitials at the final stage of the cascade simulations in both the MoNbTaVW \gls{hea} and pure W. 
Below 10~keV, the number and size of defect clusters are small in both tungsten and the HEA, making the clustered fractions of vacancies and interstitials highly susceptible to extreme values (i.e., 0\% or 100\%) in individual simulations.  
This leads to relatively large standard deviations.  
Nevertheless, the average clustered fractions still capture the overall trend.  
As the PKA energy increases, the average clustered fractions of vacancies and interstitials become more stable.
Across the entire energy range, the fraction of clustered interstitials in the \gls{hea} remains below 25\%, whereas in pure W, the fraction increases markedly with energy and approaches nearly 90\% at 150~keV.  
In contrast, the fraction of clustered vacancies in the \gls{hea} consistently exceeds that in pure W.  
It is evident that the \gls{hea} exhibits enhanced vacancy clustering and suppressed interstitial cluster formation.
This trend is also reflected, as shown in Fig.~\ref{fig:cluster-all}(b), in the size of the largest vacancy and interstitial clusters as a function of energy for both the MoNbTaVW \gls{hea} and pure W.
Overall, the number of atoms in the largest clusters increases with energy. 
In pure W, interstitial clusters larger than vacancy clusters are formed across the energy range, while the opposite phenomenon is observed in the \gls{hea}.  
Notably, large interstitial and vacancy clusters containing more than 100 atoms are still directly formed in both pure W and the \gls{hea} at a PKA energy of 150~keV.
Given that the simulations are conducted at low doses where cascade overlap is absent, the direct formation of large clusters (i.e., dislocation loops) at a high PKA energy of 150~keV is expected to play a dominant role in determining the radiation resistance. 
It is necessary to further investigate the formation and distribution of clusters at 150~keV.

\begin{figure}[H]
\centering
\includegraphics[width=\columnwidth]{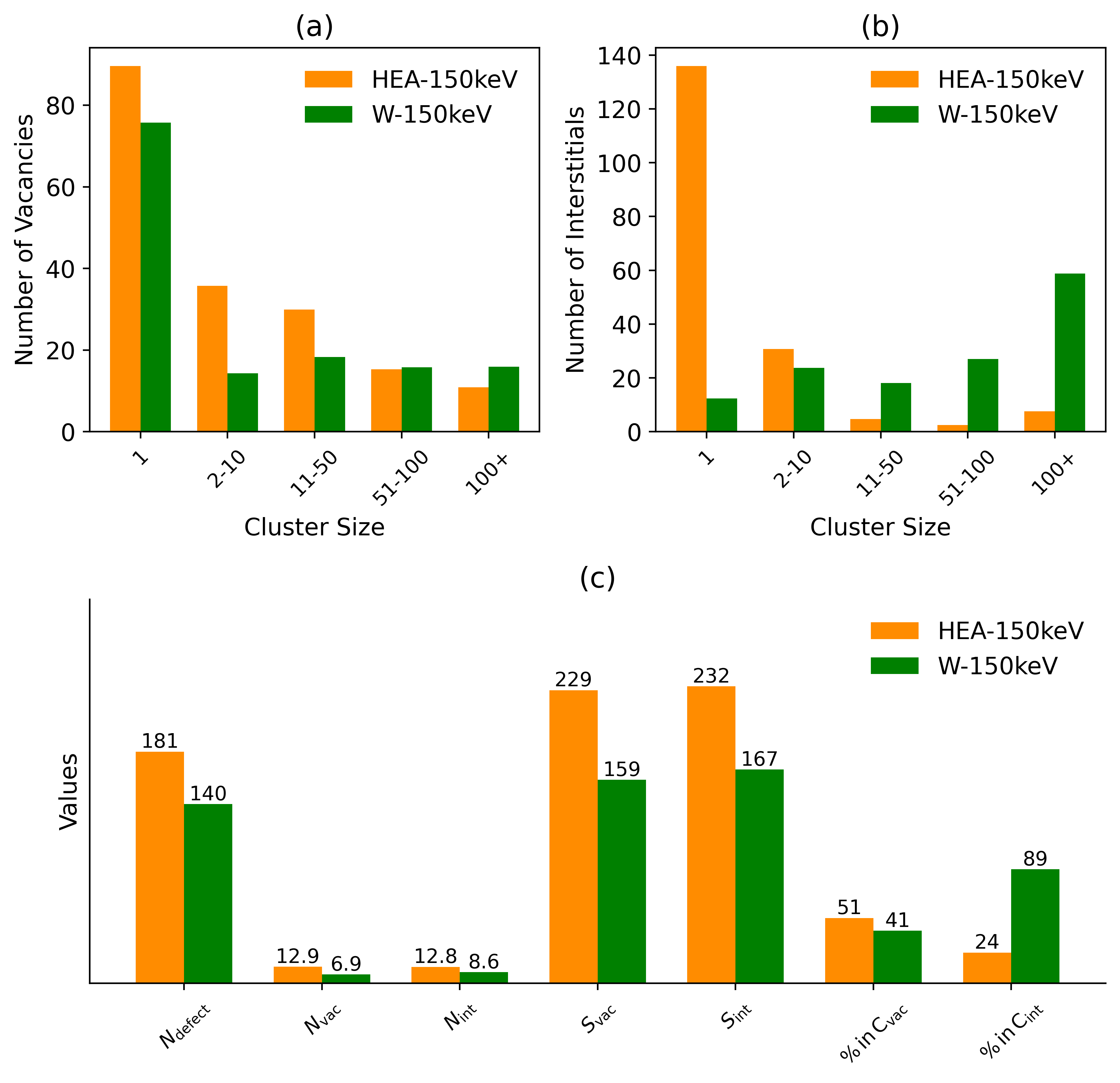}
\caption{Size distribution of (a) vacancies clusters and (b) interstitial clusters for cascades simulated in tungsten and HEAs; (c) statistical results of average number of point defects ($N_{\rm{defect}}$),  average number of vacancy and interstitial clusters ($N_{\rm{vac/int}}$), the largest vacancy and interstitial clusters ($S_{\rm{vac/int}}$) following the displacement cascade, and the percentage of vacancies and interstitials in clusters, with all clusters containing three or more vacancies and four or more interstitials.}
\label{fig:cluster}
\end{figure}

\begin{figure*}[htb]
\centering
\includegraphics[width=2\columnwidth]{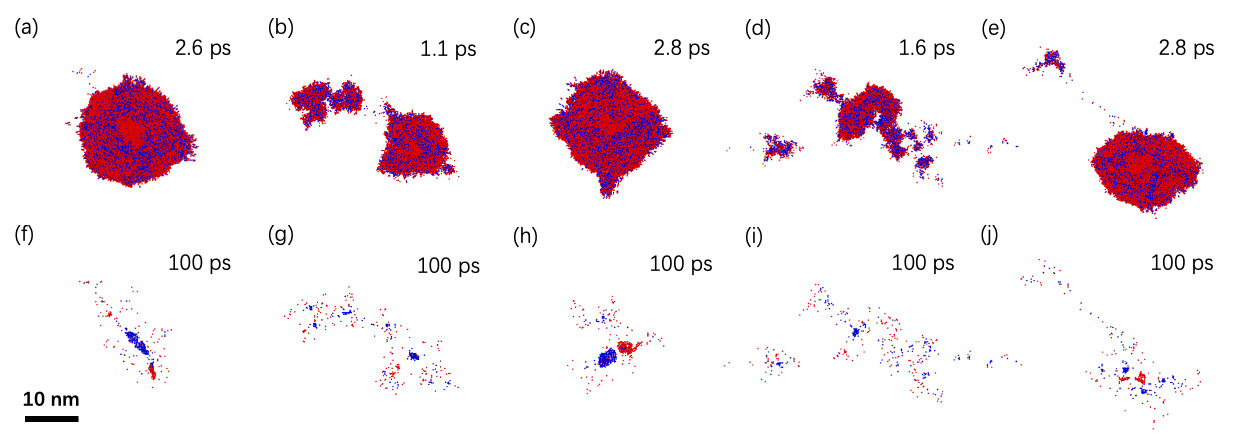}
\caption{Snapshots of cascades at the peak damage states induced by different atomic \glspl{pka}: (a) Mo, (b) Nb, (c) Ta, (d) V, (e) W. Below are images of the remaining defects in the final damage states for each type, specifically showing cascades that generated the largest interstitial clusters. Red particles represent interstitial atoms; blue particles denote vacancies.}
\label{fig:peak}
\end{figure*}

\subsection{Effect of PKA element}

Figure~\ref{fig:cluster} presents the size distributions of interstitial and vacancy clusters in pure W and the MoNbTaVW \gls{hea} at a \gls{pka} energy of 150~keV.  
At a given PKA energy, the initial velocities of different \gls{pka} types vary significantly due to their differences in atomic mass.  
To properly account for this effect, we conducted 50 cascade simulations in the MoNbTaVW \gls{hea} at 150~keV, with each element selected as the \gls{pka} 10 times.
Statistical analysis was performed on the average number of point defects, the average number of clusters, the sizes of the largest vacancy and interstitial clusters, and the percentage of vacancies and interstitials in clusters.
In MoNbTaVW \gls{hea}, vacancy clusters tend to form smaller clusters, possibly influenced by the binding energy of vacancies~\cite{Wei2024acta}, but there is an overall promotion of vacancy cluster formation.
This phenomenon is similar to that observed in the displacement cascade simulations of W-Ta-Cr-V conducted by Chen \textit{et al.} \cite{Chen2023jnm}.
However, it is evident that smaller and fewer interstitial clusters are produced in the \gls{hea} and remain more isolated instead of efficiently clustering in tungsten.
As shown in Figure~\ref{fig:cluster}(c), more \glspl{fp} are formed and more vacancy clusters are observed in the \gls{hea}, but the percentage of vacancies in clusters only slightly increases. 
In contrast, the number of interstitial clusters and the percentage of interstitials in clusters both significantly decreased in the \gls{hea}. 
Overall, compared to pure \gls{w}, the formation of defect clusters in the \gls{hea} is suppressed.

\begin{table*}[htb]
\centering \setlength{\tabcolsep}{8mm} 
\caption{Statistical results for each pka type, including the average number of point defects ($N_{\rm{defect}}$), average number of vacancy and interstitial clusters ($N_{\rm{vac/int}}$), the largest vacancy and interstitial clusters ($S_{\rm{vac/int}}$) following the displacement cascade, and the percentage of vacancies and interstitials in clusters.}
\begin{tabular}{lllllll}
\hline \hline 
type&$N_{\rm{defect}}$&$N_{\rm{vac}}$&$N_{\rm{int}}$&$S_{\rm{vac}}$&$S_{\rm{int}}$&\% in clusters\\ 
\hline 
V  & 187 & 16.9 & 14.1 & 38  & 8   & 40.5\% and 18.3\% \\
Nb & 187 & 14.5 & 13.6 & 69  & 12  & 47.5\% and 18.4\% \\
Mo & 194 & 14.5 & 13.5 & 198 & 149 & 46.2\% and 24.1\% \\
Ta & 176 & 9.4 & 11.9 & 229 & 232 & 61.4\% and 29.4\% \\
W  & 162 & 9.6 & 11.1 & 117 & 53  & 57.8\% and 27.2\% \\
\hline \hline 
\end{tabular}
\label{table:cluster-element}
\end{table*}

Significant subcascade splitting was observed, particularly in simulations with V or Nb as the \gls{pka}. 
This phenomenon occurred in all 10 simulations for these elements. 
Mo as the \gls{pka} also exhibited subcascade splitting but displayed an unfragmented peak damage state in only one of the simulations.
When W and Ta are selected as \glspl{pka}, subcascade splitting still occurred in half the instances.
At the same energies, no subcascade splitting is observed in tungsten, consistent with the subcascade splitting threshold for self-ions near 160 keV as determined by the analysis of \gls{bca} cascades \cite{De_Backer2016epl}.
Table~\ref{table:cluster-element} lists the statistical data for different types of \glspl{pka}.
Compared to the more common subcascade splitting cases, the unfragmented peak damage state leads to an increase in the percentage of vacancies within clusters, while the number of vacancy clusters decreases, resulting in fewer but larger vacancy clusters.
Large interstitial clusters are accompanied by the formation of large-sized vacancy clusters, resulting in a significant increase in the percentage of interstitials in clusters.
Figure~\ref{fig:peak} shows snapshots of cascades at the peak and final damage states, induced by different atomic \gls{pka}, which correspond to the largest interstitial clusters formed.
In subcascade splitting cases, most point defects are isolated and dispersed rather than clustered. 
Conversely, unfragmented cases exhibit defect clustering and require a longer time to reach the thermal spike.

\begin{figure}
\centering
\includegraphics[width=0.95\columnwidth]{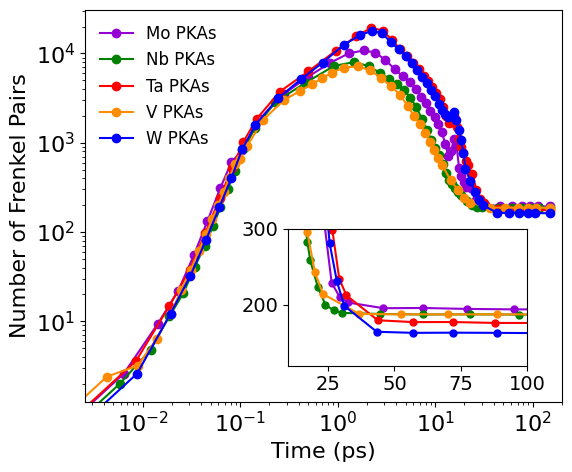}
\caption{The number of Frenkel pairs (FPs) as a function of simulation time for the MoNbTaVW \gls{hea} during cascade simulation at 150 keV with different atomic \glspl{pka}. Each point is the average of 10 independent cascade simulations, all lasting about 100 ps.}
\label{fig:spike}
\end{figure}

Figure~\ref{fig:spike} presents the cascade process induced by different atomic \glspl{pka}, accounting for the number of \glspl{fp} as a function of simulation time.
Each point is the average of 10 independent cascade simulations, all lasting about 100 ps.
Heavier atoms selected as \glspl{pka} generated more \glspl{fp} at the thermal spike, but fewer surviving \glspl{fp} after recombination.
The unfragmented heat spike resulted in the formation of large molten regions, which leading to the formation of large clusters, significantly increased the number of \glspl{fp} when cluster sizes exceeded 100 defects. 
However, when only small and medium-sized clusters were produced, it also promoted the recombination of interstitials and vacancies, ultimately resulting in a decrease in the number of \glspl{fp}.
For Mo, although the unfragmented peak damage state occurred only once, it led to formation of large clusters. 
This instance resulted in an increased average number of \glspl{fp} at both the peak and final states, as well as a higher percentage of interstitials in clusters.
For W and Ta, longer lifetime of the thermal spike promoted the recombination of interstitials and vacancies.
When Ta is selected as \glspl{pka}, the production of a large interstitial cluster exceeding 200 defects results in a higher average number of surviving \glspl{fp} than for W \glspl{pka}.
In summary, we found that cascade splitting significantly suppresses cluster formation while also hindering defect recombination, leading to the generation of more isolated and dispersed point defects.

\section{Discussion}

Setyawan \textit{et al.}~\cite{Setyawan2015jnm} reported two regions of energy dependence for defect generation in metals. 
\Gls{md} simulations were conducted with \gls{pka} energies ranging from 1 to 100 keV in tungsten, using a power-law exponent to characterize the number of Frenkel pairs produced within each region.
Setyawan \textit{et al.} proposed that the intersection of two fitted lines represents the transition energy, marking both the morphological transition of cascades and the onset of large interstitial cluster formation.
Fu \textit{et al.}~\cite{Fu2019jnm} employed the same function to investigate this dependence in pure W, 5 at.\% Re, and 10 at.\% Re alloys:
\begin{equation}
N_{\rm FP} = a(E_{\rm PKA})^b,
\end{equation}
where $N_{\rm FP}$ is the number of \glspl{fp}, $E_{\rm PKA}$ (keV) is the \gls{pka} energy, and $a$ and $b$ are fitting parameters. 
For our \gls{w} results, the parameters are $a=2.14$ and $b=0.58$ in the lower energy region, while in higher energy regions they are $a=0.16$ and $b=1.35$. 
These values are close to previous results~\cite{Setyawan2015jnm, Fu2019jnm}, though the pre-factor is slightly lower because previous studies did not consider electronic stopping.

\begin{figure}[htb]
\centering
\includegraphics[width=\columnwidth]{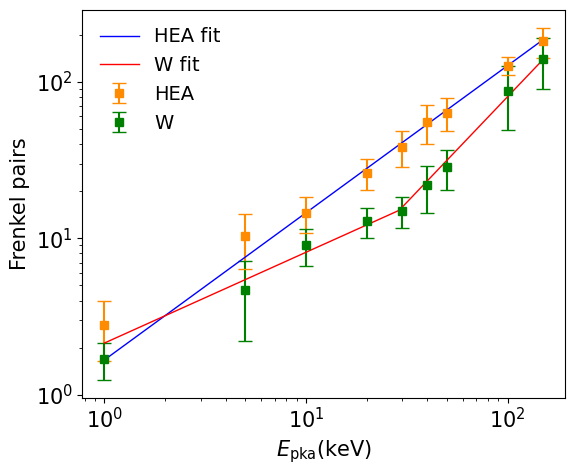}
\caption{The residual point defects and the power-law fits for W and the MoNbTaVW HEA. Error bars are given as the standard deviations.}
\label{fig:power_lawer}
\end{figure}

For the \gls{hea}, no evident difference in energy dependence is observed, with $a=1.67$ and $b=0.94$.
Combining the results of \gls{md} simulations, we attribute this phenomenon to the suppression of interstitial clusters formation caused by the chemical complexity of the alloy.
In lower energy regions, 50 cascade simulations were performed, with 10 independent runs conducted for each PKA element (Mo, Nb, Ta, V, and W) at 30, 40, and 50~keV.  
Statistical analysis of these simulations is provided in the Supplementary Materials.  
The average number of surviving \glspl{fp} exhibits an inverse correlation with \gls{pka} mass, consistent with the trends observed at 150~keV.
Relatively large vacancy clusters were direct formed, while all interstitial clusters remained smaller than 20 atoms.
This suppression of interstitial cluster formation is supported by existing studies~\cite{Gao2020acta, Wei2024acta, Chen2023jnm, Qiu2024jnm}, which have attributed it to higher defect recombination efficiency and smaller interstitial loop binding energies. 
In higher energy regions, we suggest that subcascade splitting, beginning at tens of keV \gls{pka} energies, leads to most interstitials being isolated and dispersed rather than clustered.
Especially when lighter atoms are selected as the \gls{pka}, subcascade splitting is more likely to occur, and forming large interstitial clusters directly becomes nearly impossible, even at high energies.
In the alloy, the decline in material properties is primarily due to mobile interstitial clusters reacting with each other to form more complex dislocation networks.
Compared to interstitial clusters, vacancy clusters are more difficult to move, and studies have shown that the overlap of a cascade with a vacancy-type defect decreases the number of new defects~\cite{Fellman2019jpcm}.
Therefore, the sharp reduction in interstitial clusters, driven by the combined effects of smaller interstitial loop binding energies and subcascade splitting, is key to the radiation resistance in high-entropy alloys.

Based on the extended damage model and the \gls{md} results, lighter alloying elements play a crucial role in the radiation resistance of high-entropy alloys. 
These elements increase the difficulty of atomic displacement and decrease the energy transfer cross-section, which promotes the occurrence of cascade splitting.
It is noted that large-sized interstitial clusters have been directly observed in the MoNbTaVW \gls{hea} with 150 keV \gls{pka} energies, consistent with experimental results from irradiated coarse-grained WTaCrV~\cite{Atwani2023HEAM}.
Out of fifty cascade simulations, only two interstitial clusters larger than 100 atoms were observed, indicating that the formation of such large interstitial clusters remains a rare event.
However, no signs of radiation-induced dislocation loops have been observed in nanocrystalline thin films of W-based \glspl{hea}~\cite{Atwani2019sciadv, Atwani2023nc}. 
The role of grain boundaries hence deserves further investigation.

\section{Conclusion}

We propose extended damage models for multi-component systems, and performed a systematic computational study of primary radiation damage in MoNbTaVW \glspl{hea} and pure tungsten.
The extended models demonstrated that lighter elements increase the difficulty of atomic displacement. 
An efficient machine-learned interatomic potential for the MoNbTaVW quinary system was trained, achieving computational speeds comparable to the \gls{eam} potential and enabling large-scale \gls{md} simulation.
We demonstrated its accuracy through evaluations of elastic properties, melting points, and defect energetics relevant to radiation damage. 
Additionally, \glspl{tde} in the MoNbTaVW HEA were investigated and compared with those of pure metals, highlighting compositional effects.
Using \gls{md} simulation results, we fitted the parameters of the extended \gls{arc} model to accurately predict the number of \glspl{fp}.
Further investigations into \gls{pka} types at 150 keV recoil energy reveal that the promotion of subcascade splitting is a key mechanism for enhancing radiation damage resistance in \glspl{hea}.
Specifically, subcascade splitting leads to most interstitials being isolated and dispersed rather than clustered, which significantly suppresses interstitial cluster formation and enhances radiation tolerance.
Combining the extended damage model and the \gls{md} results, it is evident that lighter alloying elements play a special role in enhancing the radiation resistance of \glspl{hea}.
This study provides critical guidance for the design of alloy compositions and enhances our insight into radiation tolerance mechanisms in high-entropy alloys.

\textbf{Data availability}

The training and validation results for the NEP model can be freely available from the Zenodo repository \url{https://doi.org/10.5281/zenodo.13948627}. Other data presented in this paper are available from the corresponding authors upon reasonable request.

\begin{acknowledgments}
The authors acknowledge funding from the National Natural Science Foundation of China (NSFC) (No. 92270001) (Y. S. \& J. L.), the Research council of Finland through grant no. 354234 (J. B.), the CNNC Science Fund for Talented Young Scholars FY222506000902 and the President Funding of CIAE YZ232602000702 (B. B.), the National Key Research and Development Program of China under grant no. 2023YFB3506704 (P. Q.), and USTB MatCom of Beijing Advanced Innovation Center for Materials Genome Engineering. 
\end{acknowledgments}

\bibliography{ref.bib}

\end{document}